\documentclass[
reprint,
superscriptaddress,
amsmath,amssymb,
aip,
floatfix,
]{revtex4-2}

\usepackage{graphicx}
\usepackage{dcolumn}
\usepackage{bm}
\usepackage{booktabs}
\usepackage{siunitx}

\usepackage{acro}
\DeclareAcronym{GAP}{short=GAP, long=Gaussian approximation potentials}
\DeclareAcronym{GP}{short=GP, long=Gaussian process, long-plural=es}
\DeclareAcronym{HetSys}{short=HetSys, long=EPSRC Centre for Doctoral Training in Modelling of Heterogeneous Systems}
\DeclareAcronym{MCMC}{short=MCMC, long=Markov chain Monte Carlo}
\DeclareAcronym{HMC-NUTS}{short=HMC-NUTS, long=Hamiltonian Monte Carlo with a no-u-turn sampler}
\DeclareAcronym{MLIP}{short=MLIP, long=machine learning interatomic potential}
\DeclareAcronym{qoi}{short=QoI, long=quantity of interest}
\DeclareAcronym{qois}{short=QoIs, long=quantities of interest}
\DeclareAcronym{UQ}{short=UQ, long=uncertainty quantification}
\DeclareAcronym{WCPM}{short=WCPM, long=Warwick Centre for Predictive Modelling}
\DeclareAcronym{SOAP}{short=SOAP, long=Smooth Overlap of Atomic Positions}
\DeclareAcronym{NAFEMS}{short=NAFEMS, long=National Agency for Finite Element Methods and Standards}
\DeclareAcronym{ACE}{short=ACE, long=atomic cluster expansion}
\DeclareAcronym{MACE}{short=MACE, long=message passing atomic cluster expansion}
\DeclareAcronym{SNAP}{short=SNAP, long=spectral neighbor analysis potential}
\DeclareAcronym{NUTS}{short=NUTS, long=No U-turn Sampler}
\DeclareAcronym{LML}{short=LML, long=linear machine learning}
\DeclareAcronym{QM/MM}{short=QM/MM, long=quantum mechanics/molecular mechanics}
\DeclareAcronym{ML}{short=ML, long=marginal likelihood}
\DeclareAcronym{LOO-CV}{short=LOO-CV, long=leave-one-out cross validation}
\DeclareAcronym{MAP}{short=MAP, long=\textit{maximum a posteriori}}
\DeclareAcronym{ANN}{short=ANN,long=artificial neural network}
\DeclareAcronym{CRPS}{short=CRPS, long=continuous ranked probability score}
\DeclareAcronym{MAE}{short=MAE, long=mean absolute error}
\DeclareAcronym{RMSE}{short=RMSE, long=root-mean-square error}
\DeclareAcronym{POPS}{short=POPS, long=pointwise optimal parameter sets}
\DeclareAcronym{MPNN}{short=MPNN, long=message passing neural network}
\DeclareAcronym{CDF}{short=CDF, long=cumulative distribution function}

\begin{document}

\title{Uncertainty quantification design principles for machine learning interatomic potentials: lessons learned from hierarchical Bayesian inference}

\author{Brennon L. Shanks}
\affiliation{Institute of Organic Chemistry and Biochemistry Prague, Czech Academy of Sciences, Prague, Czech Republic}
\author{George Simmons}
\affiliation{Warwick Centre for Predictive Modelling, School of Engineering, University of Warwick, Coventry, CV4 7AL, United Kingdom}
\author{Albert P. Bart\'ok}
\affiliation{
Department of Physics, University of Warwick, Coventry, CV4 7AL, United Kingdom}
\affiliation{Warwick Centre for Predictive Modelling, School of Engineering, University of Warwick, Coventry, CV4 7AL, United Kingdom}
\author{James R. Kermode}
\affiliation{Warwick Centre for Predictive Modelling, School of Engineering, University of Warwick, Coventry, CV4 7AL, United Kingdom}

\date{\today}

\begin{abstract}

Reliable uncertainty quantification (UQ) remains a major challenge for machine learning interatomic potentials (MLIPs). 
Here, we benchmark UQ strategies for message passing neural networks such as MACE and Gaussian process (GP)-based interatomic potentials using coupled-cluster reference data for argon as a test case. 
We develop a hierarchical GP potential combining physics-informed priors with Bayesian propagation of hyperparameter uncertainty that retains useful discrimination of prediction errors while remaining conservative rather than overconfident, with a probabilistic structure that allows the sources of predictive uncertainty to be explicitly dissected.
These results expose important distinctions between accuracy, calibration, sharpness, and error discrimination and provide practical design principles for MLIP UQ in uncertainty-critical applications.

\end{abstract}

\maketitle

\section{\label{sec:intro}Introduction}

Multiscale materials modelling has been successful in many domains but has not yet fully delivered on its promise, in large part because coupling of scales remains \emph{ad hoc}, lacking the robust propagation of uncertainties necessary to derive validated material properties at larger scales.
\Ac{MLIP} methods, in which surrogate models are trained to map atomic configurations to quantum level energy and force data, are widely used despite severe limitations in accuracy and transferability.
Recent work has therefore examined how predictions vary across ensembles of interatomic potentials with different parameterisations or training procedures.\cite{Frederiksen2004-bj,Mishra2018-vd,Tran2020-so}
Related approaches construct committees from models trained on subsets of the  data \cite{Musil2019-ef,Imbalzano2021-al} or ensembles obtained from differing random initialisations or architectures \cite{kellner_uncertainty_2024, liu_heterogeneous_2025}. 
Such approaches can provide useful empirical measures of model disagreement, but the resulting ensemble does not necessarily have a well-defined probabilistic interpretation for downstream uncertainty propagation.

The more challenging task of quantifying model-form error requires sampling over ensembles of interatomic potentials that are not constrained by a predefined functional form, allowing them to better reflect the complexity of the underlying quantum mechanics.
This is closely aligned to the broad body of recent literature that reports efforts to construct \acp{MLIP} that interpolate from a database of reference \emph{ab initio} calculations, e.g.\ widely used artificial neural network\cite{Behler2007-oq} and \ac{GAP}\cite{Bartok2010-pw} approaches, the \ac{SNAP} method~\cite{Thompson2015-zd} and related  \ac{LML}~\cite{Deres2021} approach, as well as the \ac{ACE}~\cite{Drautz2019-cq} and \ac{MACE} \cite{batatia_mace_2022} frameworks.
Data-driven models avoid many of the pitfalls of a fixed functional form, but the errors induced by overly flexible models are challenging to systematically quantify.
A range of approaches has consequently been developed for uncertainty-aware \acp{MLIP}, including \ac{POPS} regression \cite{swinburne_parameter_2025}, conformal prediction \cite{ho_flexible_2026}, and D-optimality \cite{shuang_model_2026}.
These methods address different aspects of uncertainty estimation, calibration, and active learning, but do not generally provide a hierarchical probabilistic model in which distinct sources of predictive uncertainty can be explicitly represented and propagated.

\acp{MLIP} based on \ac{GP} regression in principle provide rigorous posterior variance estimates that can address the \ac{UQ} problem.
However, this capability has been underexploited because of the tendency of \ac{GP} predicted errors to overestimate the true error, combined with the high computational cost and complexity of optimising or sampling hyperparameters.
For example, a \ac{GP}-based potential for refractory high entropy alloys~\cite{Byggmastar2021-pu} was limited to two- and three-body terms only, and then also had to be remapped from \acp{GP} to splines for computational efficiency, removing the \ac{UQ} capabilities.
Furthermore, disagreement and conflicting results reported for \ac{UQ} strategies applied to rapidly emerging foundation models suggest that a more careful and measured approach to interatomic potential \ac{UQ} is warranted. 

Here, building on prior work demonstrating \ac{UQ} with \ac{GP}-based potentials,~\cite{Musil2019-ef,Bartok2010-pw,Vandermause2020-dw} we present a step towards overcoming these limitations and developing reliable \ac{UQ} for \acp{MLIP}.
Specifically, we show that a physics-informed hierarchical \ac{GP} framework, in which \ac{GP} hyperparameter uncertainty is inferred by generalised Bayesian inference and propagated to energy and force evaluations, avoids severe overconfidence and produces uncertainties that are informative of realised prediction errors. 
We identify the most important features of a reliable uncertainty model and discuss implications for the design of \acp{MLIP} more generally.
A dataset consisting of Ar dimers and trimers computed at the CCSD(T) level is used to demonstrate the principles of our newly proposed approach.

\section{\label{sec:method}Methodology}

\subsection{Data generation}

We generated a data set of 41 Ar dimers at interatomic distances uniformly spaced between 3 and 7~\AA.
In addition, a set of 1880 Ar trimers were generated by varying the bond distances between 3 and 7~\AA{} and the bond angle between $0^\circ$-$180^\circ$.
We evaluated the \emph{ab initio} energies and forces with the ORCA software package\cite{RN178,RN204,RN205} (version 6.0.0) at the coupled cluster [CCSD(T)] level.
This ensures that dispersion interactions, which are largely responsible for attraction between the Ar atoms, are accurately modelled, including their many-body character.
Consequently, the target interaction energy of an Ar trimer configuration is truly a function of all three atomic positions, and while the interaction is dominated by pair interactions, the total energy cannot be exactly decomposed into two-body terms.

In our \emph{ab initio} calculations we used the aug-cc-pVQZ basis set\cite{woon1993a} at the \verb+TightSCF+ settings.
To minimize the basis set superposition error, we employed counterpoise corrections throughout, resulting in atomization energies as our target labels.
Due to the lack of availability of analytical gradients of the CCSD(T) energies, we carried out numerical differentiation using symmetric finite displacements of \qty{1e-4}{\angstrom} in the Cartesian directions to obtain the target forces.
 
Argon was selected because it represents perhaps the simplest nontrivial system for evaluating \ac{UQ} in \ac{MLIP}s. 
Its interactions are dominated by isotropic dispersion with negligible electronic, bonding, polarization, or chemical complexity. 
Hence, argon provides a controlled setting in which deficiencies in uncertainty estimation can be examined without simultaneously introducing substantial chemical or configurational complexity.

\subsection{Gaussian process regression}

We can model any multivariate function $f(\mathbf{x})$ using a \ac{GP}. 
This allows us to define a prior over functions that is equivalent to a multivariate normal distribution for any finite realisation of the function on a grid of $N_*$ points $X_* = [\mathbf{x}_1, \mathbf{x}_2, \ldots, \mathbf{x}_{N_*}]$ so that $\mathbf{f} = [f(\mathbf{x}_i), i=1\ldots N_*]$, i.e.
\begin{equation} \label{eq:prior-gp}
    f \sim \mathcal{GP}(\mu(\mathbf{x}), k(\mathbf{x},\mathbf{x}')) \implies \mathbf{f} \sim \mathcal{N}\left(\mathbf{\mu}(\mathbf{x}), K \right)
\end{equation}
where $\mu(\mathbf{x})$ is a specified mean function (often taken as the zero function), $k(\mathbf{x}, \mathbf{x}')$ defines a kernel for the covariance between function values at input points $\mathbf{x}$ and $\mathbf{x}'$, and $K$ is the resulting covariance matrix for all pairs of inputs, i.e. $K_{ij} = k(\mathbf{x}_i, \mathbf{x}_j)$ for $i,j=1\ldots N$.

\paragraph{Function observations}

Consider a set of $N$ noisy observations $\mathbf{y} = [y_1, y_2, \ldots y_N]$  at inputs $X = [\mathbf{x}_1, \mathbf{x}_2, \ldots \mathbf{x}_N]$ which are assumed to follow the model $f(\mathbf{x})$ on average but contaminated with additive Gaussian noise 
\begin{equation}
 y_i = f(\mathbf{x}_i) + \epsilon \; \text{where} \; \epsilon \sim \mathcal{N}(0,\, \sigma_n^2)   
\end{equation}
where $\sigma_n$ is a hyperparameter representing the noise level. Assuming additive independence, the covariance matrix for the observations is
\begin{equation} \label{eq:likelihood}
\mathrm{cov}[\mathbf{y}] = K_y = K(X, X) + \sigma_n^2 I.
\end{equation}

Conditioning the prior \ac{GP}~\eqref{eq:prior-gp} on the data leads to a new \ac{GP} representing our updated knowledge of the function, referred to as the posterior \ac{GP}. 
The posterior \ac{GP} can be used to compute the predictive mean and covariance at input points of interest $X^*$ analytically~\cite{Rasmussen2006-xl}
\begin{align}
 \bar{\mathbf{f}}(X^*) &= K(X_*,X) K_y^{-1} \mathbf{y}  \label{eq:mean} \\
 \mathrm{cov}[\mathbf{f}^*] &= K(X_*, X_*) - 
 K(X_*,X) K_y^{-1} K(X,X_*). \label{eq:cov}
\end{align}

\paragraph{Derivative observations}

The \ac{GP} framework can be extended directly to observations of derivatives of the unknown function. 
Let
\begin{equation}
    \mathbf{y}_{d} = [y_{d,1},y_{d,2},\ldots,y_{d,M}]
\end{equation}
denote $M$ derivative observations at inputs $X_d=[\mathbf{x}_{d,1},\mathbf{x}_{d,2},\ldots,\mathbf{x}_{d,M}]$.
Because differentiation is a linear operation, derivatives of a \ac{GP} remain jointly Gaussian with the function values. 
Defining the derivative observation operator $\mathcal{D}$, the relevant covariance matrices are obtained by applying $\mathcal{D}$ to the corresponding arguments of the kernel,
\begin{align}
    K_{fd}(X,X_d)
    &= \mathcal{D}_{X_d} K(X,X_d), \\
    K_{df}(X_d,X)
    &= \mathcal{D}_{X_d} K(X_d,X), \\
    K_{dd}(X_d,X_d)
    &= \mathcal{D}_{X_d}
       \mathcal{D}_{X_d'} K(X_d,X_d').
\end{align}
For force observations of a potential energy $f(\mathbf{x})$, the observation operator is $\mathcal{D}=-\nabla_{\mathbf{x}}$, since $\mathbf{F}=-\nabla_{\mathbf{x}}f$.

Assuming independent Gaussian noise with variance $\sigma_d^2$ on the derivative observations, their covariance is
\begin{equation}
    K_{y_d} = K_{dd}(X_d,X_d) + \sigma_d^2 I.
\end{equation}
Conditioning only on derivative observations gives the predictive mean and covariance of the latent function at $X_*$,
\begin{align}
    \bar{\mathbf{f}}(X_*)
    &=
    \mu(X_*)
    +
    K_{fd}(X_*,X_d)
    K_{y_d}^{-1}
    \left[
        \mathbf{y}_d-\mathcal{D}_{X_d}\mu(X_d)
    \right],
    \label{eq:derivative-mean}
    \\
    \mathrm{cov}[\mathbf{f}_*]
    &=
    K(X_*,X_*)
    -
    K_{fd}(X_*,X_d)
    K_{y_d}^{-1}
    K_{df}(X_d,X_*).
    \label{eq:derivative-cov}
\end{align}

\paragraph{Joint function and derivative observations}

Function values and their derivatives can be conditioned on simultaneously by concatenating the observation vectors,
\begin{equation}
    \mathbf{y}_{J}
    =
    \begin{bmatrix}
        \mathbf{y} \\
        \mathbf{y}_d
    \end{bmatrix}.
\end{equation}
The corresponding prior mean and covariance are
\begin{equation}
    \boldsymbol{\mu}_{J}
    =
    \begin{bmatrix}
        \mu(X) \\
        \mathcal{D}_{X_d}\mu(X_d)
    \end{bmatrix},
\end{equation}
and
\begin{equation}
    K_J
    =
    \begin{pmatrix}
        K(X,X)
        &
        K_{fd}(X,X_d)
        \\
        K_{df}(X_d,X)
        &
        K_{dd}(X_d,X_d)
    \end{pmatrix}.
\end{equation}
Assuming independent Gaussian noise for the function and derivative observations, the joint observation covariance is
\begin{equation}
    K_{y,J} = K_J +
    \begin{pmatrix}
        \sigma_n^2 I & 0 \\
        0 & \sigma_d^2 I
    \end{pmatrix}.
    \label{eq:joint-covariance}
\end{equation}
The test-training cross-covariance is
\begin{equation}
    K_{*,J} =
    \begin{bmatrix}
        K(X_*,X)
        &
        K_{fd}(X_*,X_d)
    \end{bmatrix}.
\end{equation}
The posterior predictive mean and covariance then follow directly from the standard \ac{GP} conditioning equations,
\begin{align}
    \bar{\mathbf{f}}(X_*)
    &=
    \mu(X_*)
    +
    K_{*,J}
    K_{y,J}^{-1}
    \left(
        \mathbf{y}_J-\boldsymbol{\mu}_J
    \right),
    \label{eq:joint-mean}
    \\
    \mathrm{cov}[\mathbf{f}_*]
    &=
    K(X_*,X_*)
    -
    K_{*,J}
    K_{y,J}^{-1}
    K_{*,J}^{T}.
    \label{eq:joint-cov}
\end{align}

For the interatomic potentials considered here, $\mathbf{y}$ corresponds to configuration energies and $\mathbf{y}_d$ to the associated Cartesian force components. 
Thus, a training set containing $N$ configurations contains $N$ energy observations but generally many more force observations. 
The energy+force models considered below therefore contain substantially more information than energy-only models trained on the same number of configurations.

\subsection{Physics-informed prior specification}

Physics-based knowledge can be encoded into the \ac{GP} through specification of a non-zero prior mean, which induces a posterior analogous to \eqref{eq:mean}:
\begin{align}\label{eq:nonzeromean}
 \bar{\mathbf{f}}(X^*) &= \mu(X_*) + K(X_*,X) K_y^{-1} (\mathbf{y} - \mu(X)) \\
 \mathrm{cov}[\mathbf{f}^*] &= K(X_*, X_*) - 
 K(X_*,X) K_y^{-1} K(X,X_*) \label{eq:covexpmean}.
\end{align}
It is evident that Eqs. \eqref{eq:nonzeromean} and \eqref{eq:mean} differ only by re-centering the regression problem: the training targets $\mathbf{y}$ are shifted by the prior mean $\mu(X)$, and the mean is then reintroduced at the test locations via $\mu(X_*)$. 
Consequently, the \ac{GP} is effectively trained on the residuals $\mathbf{y} - \mu(X)$, \textit{i.e.}, it models deviations from the prescribed mean function rather than the function itself.
Note that this explicit mean function formulation is equivalent to the analytical solution from O'Hagan with explict basis functions \cite{OHagan1978} if $\mu(r) = \mathbf{h}^T(r)\boldsymbol{\beta}$ with Gaussian prior weights $\boldsymbol{\beta} \sim \mathcal{N}(\boldsymbol{b}, B)$.

The consequences of reformulating the problem in this manner are twofold. 
First, we generally lose a closed-form posterior, since introducing hyperpriors over the parameters of the mean function breaks the linear–Gaussian structure required for analytic marginalization. 
Second, we gain the flexibility to specify physics-informed mean functions and place structured hyperpriors on their coefficients. 
This comes at increased computational cost, as these additional parameters must now be inferred or optimised, but allows for more controlled and physically consistent modelling of each contribution, leading to a more complete model of uncertainty.

\paragraph{Two-body prior} 

The two-body interaction is modelled as,
\begin{equation}
    V_2(r) \sim \mathcal{GP}\left(\mu_2(r;\boldsymbol{\vartheta}_2), k_2(r, r')\right)
\end{equation}
in which $\mu_2(r;\boldsymbol{\vartheta}_2)$ is the explicit two-body mean function with hyperparameters $\boldsymbol{\vartheta}_2$ and $k_2(r, r')$ is the two-body covariance function. 
 
Rather than assuming a zero-mean prior, we adopt the flexible ($\lambda$-6) Mie potential,
\begin{equation}
U(r) = \epsilon \, \frac{\lambda}{\lambda-6}\left( \frac{\lambda}{6} \right)^{\frac{6}{\lambda-6}}\left[\left( \frac{\sigma}{r} \right)^\lambda-\left( \frac{\sigma}{r} \right)^6\right]
\end{equation}
where $\lambda$, $\sigma$, and $\epsilon$ represent the rate of repulsion, effective atomic size, and dispersion energy, respectively. 
The ($\lambda$-6) Mie potential has been shown to improve thermodynamic and structural predictions in liquid argon \cite{shanks2022} and is known to accurately reproduce liquid structure \cite{shanks_bayesian_2024}.

The residual covariance is modelled with an amplitude modulated Mat\'ern kernel, which extends the nonstationary kernel of Gibbs \cite{gibbs_bayesian_1997, higdon_non-stationary_2022} to the Mat\'ern covariance form \cite{paciorek_spatial_2006}
\begin{align}
k_2(r,r') =
&\delta_2(r)\delta_2(r')
\sqrt{
\frac{2\ell_2(r)\ell_2(r')}
{\ell_2^2(r)+\ell_2^2(r')}
}
\nonumber\\
&\times
k_{\nu}\!\left(
\frac{|r-r'|}
{\sqrt{\frac{1}{2}\left[\ell_2^2(r)+\ell_2^2(r')\right]}}
\right),
\end{align}
where $\ell_2(r)$ is a local correlation length and $\delta_2(r)$ is a local residual-amplitude envelope.

In this work we use the Mat\'ern $\nu = 3/2$ form,
\begin{align}
    k_{\nu=3/2}(d) = \left(1 + \sqrt{3}d\right)\exp(-\sqrt{3}d)
\end{align}
because it produced better-conditioned derivative covariance blocks for joint inference on energies and forces.
The $\nu = 3/2$ Mat\'ern correlation is also the least smooth Mat\'ern form that supports derivative observations.
At $r = r'$, $k_2(r,r') = \delta_2^2(r)$, meaning that $\delta_2(r)$ is interpretable as a spatially varying standard deviation.

For pair potentials (even oscillatory forms encountered in liquid metals) it is atypical to observe pronounced variations over large length scales as a function of $r$. 
This motivates taking $\ell_2(r)$ as a constant. 
Under this assumption, the asymptotic behavior of the kernel at small and large $r$ is governed by $\delta_2(r)$, which we model as the sum of a logistic-decay function with learnable variance floor and sparse data terms,
\begin{align}
\delta_2(r)
&=
\delta_{\mathrm{phys}}(r)
+
\delta_{\mathrm{floor}}
\nonumber\\
&\quad+
\delta_{\mathrm{sparse}}
\left[
1-\exp\left(
-\frac{d_{\mathrm{train}}(r)}{\ell_{\mathrm{sparse}}}
\right)
\right],
\\
\delta_{\mathrm{phys}}(r)
&=
\frac{
A_2\exp[-\gamma_2(r-r_0)]
}{
1+\exp[-(\gamma_2+\kappa_2)(r-r_0)]
}.
\end{align}
Here $A_2$ is the envelope amplitude, $\gamma_2$ is the long-range decay rate, $r_0$ is the transition location, and $\kappa_2$ is the excess sigmoid sharpness. 
The physics-informed term $\delta_{\mathrm{phys}}$ has previously been used to model structure factors in neutron and X-ray scattering applications \cite{sullivan2025}.
The floor term $\delta_{\mathrm{floor}}$ prevents the residual uncertainty from collapsing to zero, while the sparse-distance term $\delta_{\mathrm{sparse}}$ increases uncertainty in pair-distance regions poorly represented in the training set. 
The quantity $d_{\mathrm{train}}(r)$ is the smooth distance from $r$ to the nearest pair distance observed in the training configurations.

Lognormal priors are placed on the positive scale parameters $\ell_2$, $A_2$, $\gamma_2$, $\kappa_2$, $\delta_{\mathrm{floor}}$, $\delta_{\mathrm{sparse}}$, and $\ell_{\mathrm{sparse}}$, while $r_0$ is assigned a truncated normal prior on the physically relevant distance interval.
This enforces positive covariance scales and decay lengths, with the effective sigmoid sharpness $\gamma_2+\kappa_2$ constrained to be larger than the long-range decay rate $\gamma_2$. 


\subsection{Hyperparameter inference}

For the Bayesian interpretation of a kernel model to be meaningful, attention must be paid to the choice of hyperparameters $\boldsymbol{\theta}$, defined here as the set of parameters specifying the prior mean, residual covariance, and observation-noise model. 
For the two-body Mie + Gibbs-Matern model used in this work,
\begin{align}
\boldsymbol{\theta}
=
\big[
&\underbrace{\sigma,\epsilon,\lambda}_{\text{prior mean}},
\nonumber\\
&\underbrace{
\ell_2,A_2,\gamma_2,r_0,\kappa_2,
\delta_{\mathrm{floor}},
\delta_{\mathrm{sparse}},
\ell_{\mathrm{sparse}}
}_{\text{2-body covariance}},
\nonumber\\
&\underbrace{\sigma_E,\sigma_F}_{\text{noise}}
\big].
\end{align}
Here $(\sigma,\epsilon, \lambda)$ define the ($\lambda-6$) Mie prior mean, while the covariance parameters control the Mat\'ern correlation length, amplitude envelope, baseline floor, and sparse-region uncertainty boost. 
The terms $\sigma_E$ and $\sigma_F$ denote the energy and force observation-noise scales, respectively.

Up until now hyperparameters have often been set heuristically by \ac{MLIP} practitioners for reasons of simplicity and computational expediency.
However, in generalised Bayesian inference the posterior for a hierarchical probability model --- such as a \ac{GP} with uncertain hyperparameters --- is given by the generalised posterior
\begin{eqnarray}
    p(\boldsymbol{\theta}|\mathcal{D}) = \frac{\mathcal{L}(\mathcal{D}|\boldsymbol{\theta}) \;p(\boldsymbol{\theta})}{\int \mathcal{L}(\mathcal{D}|\boldsymbol{\theta}) \; p(\boldsymbol{\theta}) \; d \boldsymbol{\theta}}
\end{eqnarray}
where $\mathcal{L}$ is the generalised \textit{likelihood} and $p(\boldsymbol{\theta})$ the \textit{hyperprior} distribution over hyperparameters. 
For conventional Bayesian inference, $\mathcal{L}(\mathcal{D}|\boldsymbol{\theta})=p(\boldsymbol{y}|\boldsymbol{\theta})$ and the usual Bayesian posterior is recovered.
The denominator is a normalisation constant of the generalised hyperposterior, analogous to the partition function in statistical mechanics, which we refer to as the generalised \textit{marginal evidence}.

\paragraph{Hyperpriors} 

In this work, hyperparameters were assigned positive-support hyperpriors, primarily log-normal distributions, while the Mie potential repulsive exponent was assigned a truncated normal distribution.
Default hyperpriors and a hyperprior sensitivity analysis are reported in the Appendix~\ref{appendix:hypprior}.

\paragraph{Likelihood objectives}

The \ac{GP} likelihood objective was either the marginal likelihood, which has previously been reported to improve predicted variances for \ac{GP}-based interatomic potentials\cite{Vandermause2020-dw}, or the \ac{LOO-CV} objective, which has the advantage that it does not depend on the validity of modelling assumptions.
The generalised hyperposterior was sampled up to the marginal evidence normalisation constant, with normalisation performed \textit{post hoc}.

The marginal likelihood is the probability of the observed data given the inputs, and can be computed by integrating the likelihood multiplied by the prior (i.e. marginalising over the function values $\mathbf{f})$), leading to
\begin{equation}
    p(\mathbf{y}|X,\boldsymbol\theta) = \int p(\mathbf{y}|\mathbf{f}, X,\boldsymbol\theta) p(\mathbf{f}|X,\boldsymbol\theta) d\mathbf{f}
\end{equation}
For a zero mean \ac{GP} the prior and likelihood are both Gaussian, so the integral is analytic and yields
\begin{equation} \label{eq:marginal-likelihood}
    \log p(\mathbf{y}|X,\boldsymbol\theta) = -\frac12 \mathbf{y}^TK_y^{-1}\mathbf{y} - \frac12 \log |K_y| - \frac{N}{2} \log 2\pi 
    \textrm{,}
\end{equation}
which is simply the probability density function of the observations $\mathbf{y} \sim \mathcal{N}(\mathbf{0}, K_y)$. 
The three terms can be interpreted as a data fit term, a complexity penalty and a normalisation constant, respectively.

The \ac{LOO-CV} approach is based on the idea of leaving out each piece of training data in turn. 
The predictive log probability of the data when leaving out training case $i$ is
\begin{equation}
    \log p(y_i|X,\mathbf{y}_{-i},\boldsymbol{\theta}) = -\frac12 \log \sigma_i^2 - \frac{(y_i - \mu_i)^2}{2\sigma_i^2} - \frac12 \log 2\pi
\end{equation}
where $\mathbf{y}_{-i}$ means all observations except for number $i$, and the predictive mean $\mu_i$ and variance $\sigma_i^2$ are given by \eqref{eq:nonzeromean} and \eqref{eq:covexpmean} using reduced training sets $\{X_{-i}, \mathbf{y}_{-i}\}$. 
From this result we can write down the \ac{LOO-CV} log predictive probability as
\begin{equation}
    L_{LOO}(X,\mathbf{y},\boldsymbol\theta) = \sum_{i=1}^N  \log p(y_i|X,\mathbf{y}_{-i},\boldsymbol{\theta}) \label{eq:loo-cv-likelihood}
\end{equation}
which can be computed efficiently by noting that we require a series of inverses of the full $K_y$, each with a row and column deleted~\cite{Sundararajan2001-ft}.
This leads to simplified expressions for the \ac{LOO-CV} predictive mean and variance of the form
\begin{align}
    \mu_i &= y_i - [K_y^{-1}\mathbf{y}]_i/[K_y^{-1}]_{ii} \\
    \sigma_i^2 &= 1/[K_y^{-1}]_{ii}
\end{align}
and can thus be computed with a computational cost that is dominated by a single inversion of $K_y$.
Since the \ac{LOO-CV} log predictive density is not a proper likelihood, the resulting generalised hyperposterior is not a Bayesian posterior. 
Nevertheless, previous work has demonstrated that generalised hyperposteriors constructed from the \ac{LOO-CV} objective provide substantially improved \ac{UQ} for free energy calculations \cite{skorna_2026_freeenergy} and force field training based on \acp{GP} \cite{shanks_accelerated_2024, kostal_bayesian_2026}.

\begin{table*}[ht!]
\centering
\caption{Uncertainty quantification strategies benchmarked in this work.}
\label{tab:uq}
\begin{tabular}{lllll}
\toprule
Method & Source of uncertainty & Predictive form & Bayesian & Ref. \\
\midrule
Physics-informed \ac{GP} & Hyperparameter uncertainty propagation & Gaussian mixture & Yes & this work \\
MACE multi-foundation committee & Foundation backbones & Gaussian approx. & No & \cite{Bilbrey2025} \\
MACE seed committee & Random initialization / training & Gaussian approx. & No & \cite{Schran2020} \\
\bottomrule
\end{tabular}
\end{table*}

\paragraph{Hamiltonian Monte Carlo implementation}

Generalised hyperposterior sampling was performed using \ac{HMC-NUTS}~\cite{hoffman2014nuts} as implemented in \texttt{Turing.jl}~\cite{fjelde_turing_2025}. 
At each proposed parameter state, the explicit mean was evaluated, the \ac{GP} covariance matrix assembled, and target log density computed by Cholesky factorization of $\boldsymbol{K}_{\boldsymbol{\eta}}+\sigma_n^2\boldsymbol{I}$.
The log generalised hyperposterior was defined as the sum of the chosen \ac{GP} objective - either the log \ac{LOO-CV} predictive density or log marginal likelihood - and the log hyperprior.

The calculations used dispersed chain initializations to reduce sensitivity to a single starting point. 
The default production settings in the current implementation use 500 adaptation steps and 1000 retained samples per chain for the squared exponential kernel, and 500 retained samples per chain for the Gibbs kernel, with four independent chains unless otherwise specified. 
Sampling quality was assessed using standard MCMC diagnostics, including trace plots, marginal posterior histograms, effective sample size, $\hat{R}$, and divergent-transition counts.

\paragraph{Hyperposterior uncertainty propagation}

For a fixed hyperparameter state $\boldsymbol{\theta}$, the \ac{GP} predictive distribution at test structures $\boldsymbol{X}_*$ is Gaussian,
\begin{equation}
    p(\boldsymbol{f}_* \mid \boldsymbol{y}, \boldsymbol{\theta})
    =
    \mathcal{N}
    \left(
        \boldsymbol{\mu}_{\boldsymbol{\theta}},
        \boldsymbol{\Sigma}_{\boldsymbol{\theta}}
    \right).
\end{equation}
The fully Bayesian predictive distribution is obtained by marginalizing over the sampled hyperposterior,
\begin{equation}
    p(\boldsymbol{f}_* \mid \boldsymbol{y})
    =
    \int
    p(\boldsymbol{f}_* \mid \boldsymbol{y}, \boldsymbol{\theta})
    p(\boldsymbol{\theta}\mid\boldsymbol{y})
    \,d\boldsymbol{\theta},
\end{equation}
which gives a mixture of Gaussian predictive distributions. 
Given posterior samples $\{\boldsymbol{\theta}^{(m)}\}_{m=1}^M$, the predictive mean and variance are estimated by
\begin{equation}
    \bar{\boldsymbol{\mu}}
    \approx
    \frac{1}{M}
    \sum_{m=1}^{M}
    \boldsymbol{\mu}_{\boldsymbol{\theta}^{(m)}},
\end{equation}
and
\begin{equation}
    \bar{\boldsymbol{\Sigma}}
    \approx
    \frac{1}{M}
    \sum_{m=1}^{M}
    \boldsymbol{\Sigma}_{\boldsymbol{\theta}^{(m)}}
    +
    \frac{1}{M}
    \sum_{m=1}^{M}
    \left(
        \boldsymbol{\mu}_{\boldsymbol{\theta}^{(m)}}-\bar{\boldsymbol{\mu}}
    \right)
    \left(
        \boldsymbol{\mu}_{\boldsymbol{\theta}^{(m)}}-\bar{\boldsymbol{\mu}}
    \right)^{\top}.
\end{equation}
The first term is the average conditional \ac{GP} predictive uncertainty at fixed hyperparameters, while the second term captures uncertainty due to variation in the sampled mean and kernel hyperparameters. 
In the reported diagnostics, this mixture approximation is used to calculate \ac{RMSE}, \ac{CRPS}, and realised versus predicted error parity plots.

\paragraph{Continuous ranked probability score}: \ac{CRPS} measures the quality of a predictive probability distribution relative to an observed value.
For a predictive \ac{CDF} $F(x)$ and scalar observation $y$, it is defined as
\begin{equation}
    \mathrm{CRPS}(F,y)
    =
    \int_{-\infty}^{\infty}
    \left[F(x)-\mathbf{1}(x\geq y)\right]^2\,dx,
\end{equation}
where $\mathbf{1}(x\geq y)$ is the empirical \ac{CDF} associated with the observation.
For $N$ observations, the mean \ac{CRPS} reported in this work is
\begin{equation}
    \overline{\mathrm{CRPS}}
    =
    \frac{1}{N}\sum_{i=1}^{N}
    \mathrm{CRPS}(F_i,y_i).
\end{equation}
\ac{CRPS} is a proper scoring rule that rewards predictive distributions that are both accurate and appropriately dispersed. 
In expectation, it is minimized when the predictive distribution $F$ equals the true data-generating distribution $G$.
The \ac{CRPS} has the same units as the predicted quantity (meV atom$^{-1}$ in this work), with lower values indicating better probabilistic predictions.
\ac{RMSE} and parity plots are described in Appendix~\ref{appendix:eval}.

\subsection{Model and uncertainty quantification benchmarks}

\paragraph{Model potentials} The hierarchical \ac{GP} potential described above, and fine-tuned variants of the MACE-MH-0, MACE-MH-1, MACE-MPA-0 and MACE-OMAT-0 foundation models, were benchmarked on the argon cluster data.
The \ac{GP} models use two-body pair-distance descriptors throughout, with the Mie potential as the prior mean and the \ac{GP} modelling the residual energy; zero-mean variants appear only in the prior-mean comparison below.

For neural-network baselines, we fine-tuned pretrained MACE foundation models.
Multi-head checkpoints were fine-tuned from a single head (MACE-MH-0 and MACE-MH-1, using their MATPES r2SCAN and OMOL heads, respectively), while single-head checkpoints were fine-tuned directly (MACE-MPA-0 and MACE-OMAT-0)~\cite{Batatia2025-qv}.
All models were trained on the same nested data-ablation splits: training sets of 20, 40, 60 and 100 configurations drawn from the 1921 dimer and trimer configurations, with the remaining 1821 configurations held out as a fixed test set on which every energy metric below is evaluated (forces are available, and used for training, only on the 100 training configurations).

\begin{figure*}[ht!]
    \centering
    \includegraphics[width=\textwidth]{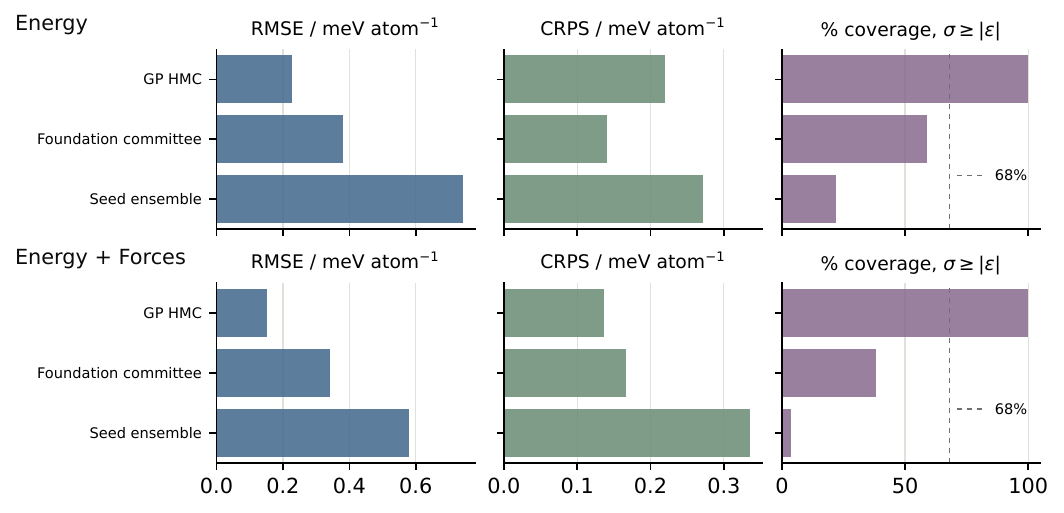}
    \caption{\ac{MLIP} training metrics for all tested methods on energy only training set (top row) and an energy + force training set (bottom row) reporting \ac{RMSE}, \ac{CRPS}, and empirical coverage.}
    \label{fig:comparison}
\end{figure*}

\paragraph{Uncertainty quantification} Uncertainty estimates from \ac{GP} posterior variances, optimised \ac{GP} hyperparameters, hyperposterior sampling, MACE foundation model committees, and MACE seed ensembles were computed. 
For fixed-hyperparameter \ac{GP} models, predictive uncertainty was given by the conditional \ac{GP} posterior variance. 
For optimised \ac{GP} models, hyperparameters were selected by either marginal likelihood or leave-one-out cross-validation. 
For the explicit-mean \ac{GP}, \ac{HMC-NUTS} was used to sample the mean, kernel, and noise hyperparameters, producing a mixture predictive distribution.

For MACE models, uncertainty was estimated using disagreement across independently trained seeds, or by fine-tuning with the same data starting from four different pretrained foundation backbones (MH-0 MATPES r2SCAN head, MH-1 OMOL head, MPA-0, OMAT-0). 

\section{\label{sec:results}Results}

\subsection{Benchmarking UQ strategies for \ac{MLIP}s}

Figure~\ref{fig:comparison} compares \ac{RMSE}, \ac{CRPS}, and empirical coverage for the hierarchical \ac{GP} and \ac{MACE} \ac{UQ} strategies at 100 training configurations for energy-only and energy+force training sets.
The energy-only hierarchical \ac{GP} predicts with an \ac{RMSE} of 0.23 meV atom$^{-1}$ and \ac{CRPS} of 0.22 meV atom$^{-1}$, with a conservative 100.0\% coverage. 
With force information, the \ac{GP} improves to an \ac{RMSE} of 0.15 meV atom$^{-1}$ and \ac{CRPS} of 0.14 meV atom$^{-1}$, again with 100.0\% coverage. 
Thus, although the hierarchical \ac{GP} pays a clear low-data penalty, it becomes more accurate while retaining conservative predictive intervals as the training set grows.

\begin{figure*}
    \centering
    \includegraphics[width=\textwidth]{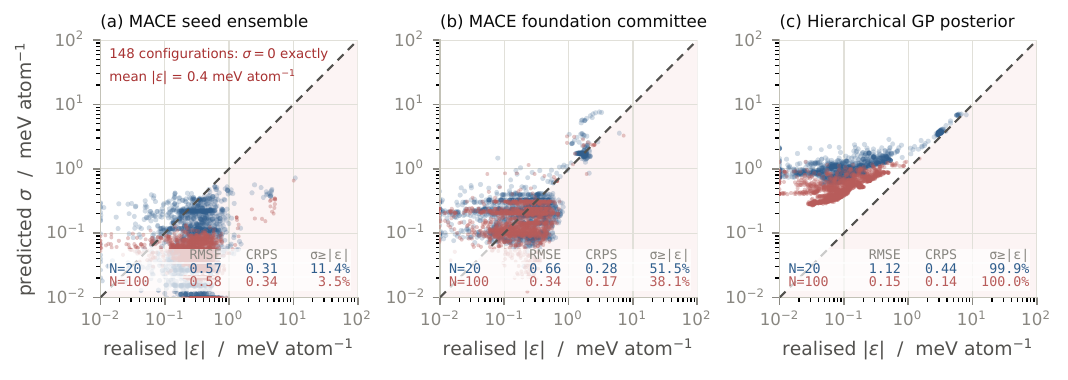}
    \caption{Energy+force trained log-log parity plots for (a) \ac{MACE} seed ensemble (5 seeds fine-tuned from the MACE-MH-0 backbone), (b) \ac{MACE} foundation model committee (comprising MACE-MH-0 MATPES r2scan head, MH-1 OMOL head, MPA-0 and OMAT-0), and (c) the hierarchical GP developed in this work. Summary statistics (\ac{CRPS}, \ac{RMSE}, and coverage) for 20 and 100 training configurations are listed in the legend.}
    \label{fig:parity}
\end{figure*}

The \ac{MACE} ensemble methods show a different tradeoff. 
The foundation committee achieves strong accuracy, particularly with energy+force information, where it reaches an \ac{RMSE} of 0.34 meV atom$^{-1}$ and \ac{CRPS} of 0.17 meV atom$^{-1}$. 
Nevertheless, its coverage is only 38.1\%, indicating strong overconfidence. 
The \ac{MACE} seed ensemble is even more overconfident, with coverage of 21.7\% in the energy-only setting and 3.5\% in the energy+force-trained setting.

Clearly, we can see a sharp distinction between accuracy and the quality of the associated uncertainty estimates in terms of \ac{CRPS} and coverage. 
\ac{MACE}-based ensembles can achieve low \ac{RMSE} and \ac{CRPS}, in this case particularly for the foundation model committee, but their low empirical coverages still indicate that they are often strongly overconfident. 
On the other hand, the hierarchical \ac{GP} model is the most accurate and achieves the best \ac{CRPS} on the energy+force-training set, but is also too conservative. 

Parity plots comparing the realised and predicted errors can give insight into evaluation metrics compared above.
Log-log plots of realised versus predicted error for the fine-tuned \ac{MACE} five-member seed ensemble, \ac{MACE} foundation backbone four-member committee, and our hierarchical \ac{GP}-based potential are shown in Figure~\ref{fig:parity}. 
Across the force-trained parity plots, the hierarchical \ac{GP} retained a positive correlation between predicted uncertainty and realized error, with Spearman coefficients of $\rho=0.62$ and $\rho=0.65$ for $N=20$ and $N=100$, respectively.
The \ac{MACE} ensemble methods showed weaker error discrimination.
For the seed ensemble, the correlation was nearly absent at $N=20$ ($\rho=0.06$) and became weakly negative at $N=100$ ($\rho=-0.18$).
The foundation committee performed better, but still showed only modest discrimination, with $\rho=0.38$ at $N=20$ and $\rho=0.21$ at $N=100$.
Thus, while the \ac{MACE} ensembles can give competitive point predictions, their ensemble spread is less reliable as a ranking signal for identifying configurations with large errors.
This behaviour is important for active learning applications, where uncertainty estimates are expected to prioritize configurations requiring additional electronic structure calculations.

The principal limitation of the hierarchical \ac{GP}, as indicated by our experiments, is its conservativeness, \textit{i.e.} its tendency to overestimate the error. 
However, looking at Figure~\ref{fig:parity}(c), we can see that much of this overcoverage occurs in the regime of low realised error, in which the predicted error asymptotically approaches a base level of uncertainty.
With increasing realised error, the predicted and realised error begin to correlate with near parity.
This result is strikingly observed in the 20-configuration case.
Moving from 20- to 100-configurations, the asymptotic tail shifts to lower predicted errors, but still maintains the same qualitative behavior.
Why does the hierarchical \ac{GP} uncertainty look this way?
The primary reason is that hierarchical Bayesian inference naturally estimates the uncertainty arising from all terms in the provided model form.
This means that, despite being able to very accurately predict unseen test data, it cannot be as confident because there is not enough information in the training set to narrow its hyperposterior distribution.

\paragraph{Variance decomposition of the hierarchical \ac{GP}}

\begin{figure*}
    \centering
    \includegraphics{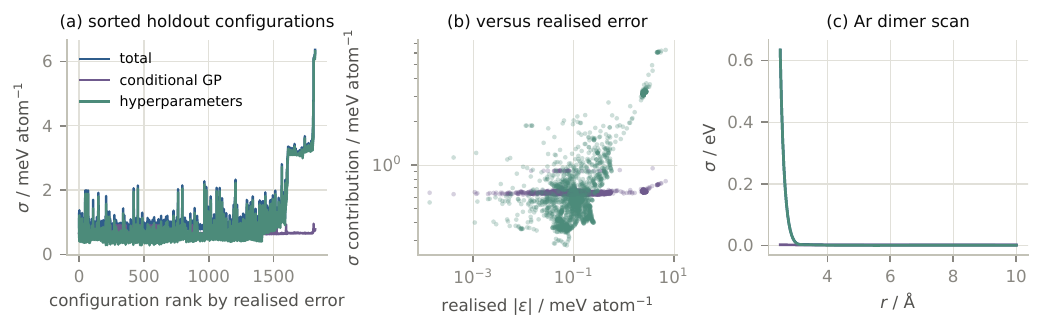}
    \caption{Decomposition of \ac{GP} predictive standard deviation into conditional and hyperparameter components.}
    \label{fig:vardecomp}
\end{figure*}

We can interrogate the hierarchical \ac{GP} predicted errors further by decomposing the posterior predictive variance into its conditional \ac{GP} and hyperparameter contributions,
\begin{align}
\mathrm{Var}(f_* \mid \mathcal{D})
&=
\mathbb{E}_{\boldsymbol{\theta}\mid\mathcal{D}}
\left[
\mathrm{Var}(f_* \mid \boldsymbol{\theta},\mathcal{D})
\right]
\nonumber\\
&\quad+
\mathrm{Var}_{\boldsymbol{\theta}\mid\mathcal{D}}
\left[
\mathbb{E}(f_* \mid \boldsymbol{\theta},\mathcal{D})
\right].
\end{align}
Here, the first term measures the usual \ac{GP} posterior uncertainty at fixed hyperparameters, while the second term measures the spread in posterior mean predictions induced by uncertainty in the prior-mean, kernel, and noise hyperparameters.
Contributions from each predictive variance term are shown in Figure~\ref{fig:vardecomp} for the 20-configuration energy+force hierarchical \ac{GP} with Mie prior mean and nonstationary Mat\'ern kernel.
The two terms contributed roughly equally when averaged uniformly over the holdout set: 48.3\% of the variance came from the conditional \ac{GP} term and 51.7\% from hyperparameter uncertainty.
However, we can see in Figure~\ref{fig:vardecomp}(b) that the behavior of each type of error contributes to the total predicted error in fundamentally different ways.
Specifically, the hyperparameter uncertainty as modelled by the hyperposterior improved the error estimation at high realised errors, whereas the conditional \ac{GP} variance term tended to establish the low-error asymptotic tail.
When weighted by the total predictive variance, the hyperparameter term therefore accounted for 81.1\% of the variance.
Thus, the broadest predictive intervals are dominated by uncertainty over plausible hyperparameter settings rather than by the conditional \ac{GP} variance alone, while the conservative tail is dominated by the conditional \ac{GP} variance.

This supports the interpretation that the hierarchical \ac{GP} is not assigning large uncertainty at large realised error because individual test configurations are locally far from the training set. 
Instead, the dominant contribution in the high-variance regime comes from variation among plausible posterior models. 
Consider that this observation also helps explain why ensemble methods become increasingly overconfident at large realised errors. 
Such approaches attempt to represent model uncertainty indirectly through variation in random initialisation, training trajectory, data selection, or model architecture. 
However, the resulting empirical ensemble does not constitute a systematic marginalisation over model or hyperparameter uncertainty and therefore samples only a narrow subset of plausible predictive models. 
Consequently, ensemble disagreement can substantially underestimate epistemic uncertainty, particularly for configurations outside the well-constrained training regime.

\subsection{Lessons for Gaussian process based \ac{MLIP}s}

The \ac{UQ} capabilities of \acp{GP} are well documented and it is not surprising that they are among the best performing strategies for \acp{MLIP} in our benchmark. 
However, not all \acp{GP} are created equal.
For example, one can build various knowledge into the \ac{GP} prior, such as physics-informed prior means and kernels, hyperparameter uncertainty propagation, different likelihood (or pseudo-likelihood) objectives, and noise models, among others, which can change the error estimation capability of the \ac{GP}.
Here we show a series of tests with different \ac{GP} variants to determine the most important features that a \ac{GP} needs to provide reliable \ac{UQ}, and also where we could potentially reduce computational expense in \ac{MLIP} design.

\subsubsection{Introducing richer physics-informed priors}

We first examined how the choice of prior mean and two-body covariance affects the hierarchical \ac{GP}. 
Four models were compared using the same 20-configuration energy+force training set: a zero-mean \ac{GP} with a squared-exponential kernel, a zero-mean \ac{GP} with a nonstationary Mat\'ern kernel, a Mie-prior \ac{GP} with a squared-exponential kernel, and a Mie-prior \ac{GP} with a nonstationary Mat\'ern kernel. 
The Mie prior mean encodes the expected short-range repulsion and long-range attraction of the Ar pair interaction, while the nonstationary Mat\'ern kernel provides a nonstationary amplitude modulation with a rougher Mat\'ern correlation.

The results show that the physics-informed prior mean is essential in the low-data regime. 
With a squared-exponential kernel, replacing the zero mean by the Mie prior reduced the \ac{RMSE} from 3.22 to 0.74 meV atom$^{-1}$ and the \ac{CRPS} from 1.16 to 0.42 meV atom$^{-1}$. 
The same effect was even more pronounced for the nonstationary Mat\'ern kernel: the zero-mean model gave an \ac{RMSE} of 5.13 meV atom$^{-1}$ and a \ac{CRPS} of 1.16 meV atom$^{-1}$, whereas the Mie-prior model reduced these to 1.10 and 0.44 meV atom$^{-1}$, respectively. 
Coverage also changed from approximately 87\% for both zero-mean models to 99\% for the Mie squared-exponential and nonstationary Mat\'ern model.

These trends indicate that the \ac{GP} covariance alone is insufficient to recover the correct interaction shape from only 20 configurations. 
The Mie prior mean supplies the correct qualitative physics, allowing the \ac{GP} to model residual corrections rather than the full interaction energy. 
This construction is closely related to the established practice of $\Delta$-learning, in which a machine-learning model is trained on the difference between a lower-level baseline and a higher-level reference~\cite{Ramakrishnan2015}.
Such an approach is particularly advantageous when the residual is at least as smooth and less complex than the target function itself.
Once the Mie mean is in place the choice of kernel matters much less: the squared-exponential kernel gives the lower \ac{RMSE} (0.74 versus 1.10 meV atom$^{-1}$) with essentially the same \ac{CRPS} and coverage, and the difference is comparable to the run-to-run variation between independent \ac{HMC-NUTS} chains of the same model (1.03--1.15 meV atom$^{-1}$ across three repeats of the Mie nonstationary Mat\'ern model).
We nevertheless retain the nonstationary Mat\'ern kernel as the working model, for the reasons given in Section~\ref{sec:method}: its $\nu=3/2$ correlation gives well-conditioned derivative covariance blocks for joint energy-force inference, and its amplitude envelope returns the posterior to the prior mean outside the region covered by training pair distances.

For Ar, the success of a Mie prior is not surprising, since accurate two-body empirical potentials are already available for this comparatively simple system.
The broader lesson is therefore not that a Mie potential should generally be used as the two-body prior, but that an established and well-benchmarked empirical potential can provide an informative baseline around which an \ac{MLIP} learns the remaining corrections.
In the present Bayesian formulation, this extends the $\Delta$-learning principle by treating the empirical potential as the prior mean of the \ac{GP}, such that uncertainty in deviations from the baseline is represented explicitly.

\begin{figure}
    \centering
    \includegraphics[width=\columnwidth]{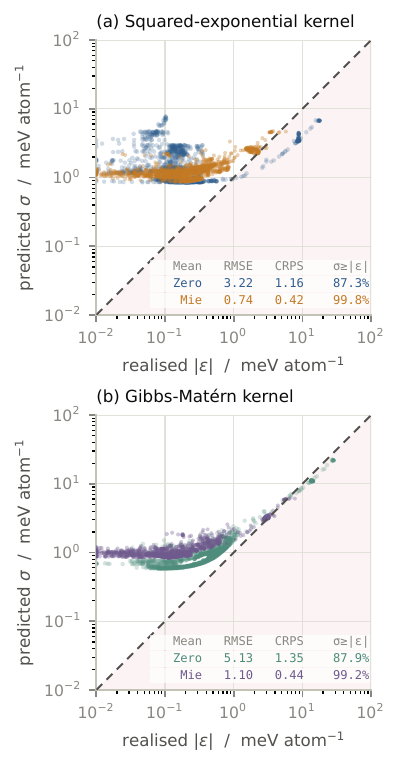}
    \caption{A comparison of \ac{GP} prior means with different kernels. Top: Zero and physics-informed ($\lambda$-6) Mie prior means with the squared exponential kernel. Bottom: Zero and physics-informed ($\lambda$-6) Mie prior means with the Gibbs-Mat\'ern kernel.}
    \label{fig:priors}
\end{figure}

\subsubsection{Importance of hyperparameter uncertainty propagation}

\begin{figure}
    \centering
    \includegraphics[width=\columnwidth]{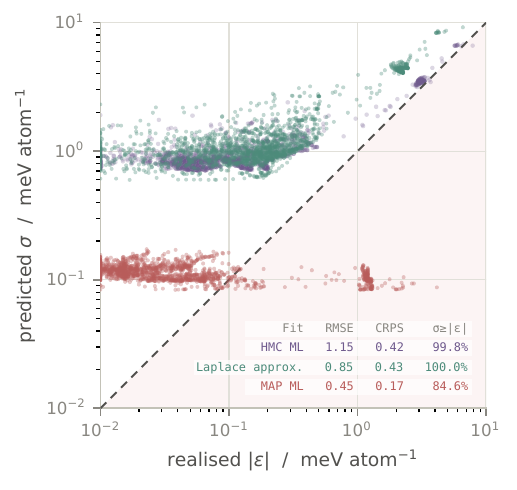}
    \caption{Comparing three hyperparameter uncertainty propagation strategies, (i) hyperposterior uncertainty propagation with \ac{HMC-NUTS}, (ii) Laplace approximation, and (iii) \ac{MAP} estimation.}
    \label{fig:uncprop}
\end{figure}

To isolate the role of hyperparameter uncertainty, we compared hyperparameter marginalised \ac{GP} predictions with the Laplace approximation (which approximates the hyperposterior as a multi-dimensional Gaussian distribution, here constructed in the unconstrained log-transformed hyperparameter space, at the mode of the transformed density and with the Hessian obtained by automatic differentiation) and \ac{MAP} hyperparameter estimates.
\ac{MAP} inference substantially improved point prediction accuracy, reducing the \ac{RMSE} and \ac{CRPS} by a factor of $\sim$2.5, while producing a less conservative empirical coverage of 84.6\%. 
However, this improvement came at the expense of uncertainty–error discrimination, as shown in Figure \ref{fig:uncprop}.
The Spearman rank correlation between predictive uncertainty and realized error decreased from $\rho$=0.50 in the hierarchical \ac{GP} case to $\rho$=-0.44, indicating that the resulting uncertainties contained little information about where the model was likely to fail. 
Indeed, a negative correlation between true and predicted errors is alarming, but not surprising.
As we demonstrated before, the hyperposterior uncertainty propagation is what distinctly enabled the hierarchical \ac{GP} to track realised errors in the high-variance regime.
Thus, although \ac{MAP} estimates provide more accurate point predictions, they largely eliminate the epistemic uncertainty represented by the hierarchical model and no longer rank errors effectively.

Notably, using the \ac{MAP} estimate in tandem with a Gaussian approximation of the hyperposterior, the so-called Laplace approximation, significantly improves the correlation between true and predicted errors, with $\rho$=0.57.
The fact that the Laplace approximation also improved \ac{RMSE} with effectively no change in \ac{CRPS} compared to the hierarchical \ac{GP} is a good indicator that this may be an effective strategy to endow \acp{MLIP} with hyperparameter \ac{UQ} without running costly \ac{HMC-NUTS} sampling. 
Of course, there is always a risk that the hyperposterior is non-Gaussian, which will cause the Laplace approximation to break down, and may explain why the hierarchical \ac{GP} achieves better calibration at high realised errors (as seen in Figure~\ref{fig:uncprop}).
The result is also sensitive to how the Gaussian is constructed: a Laplace approximation expanded instead around the \ac{MAP} estimate in the original (constrained) hyperparameter coordinates, with a finite-difference Hessian, behaved much more like the \ac{MAP} model itself (\ac{RMSE} 0.45 meV atom$^{-1}$, \ac{CRPS} 0.17 meV atom$^{-1}$, coverage 86.5\%, $\rho$=0.39), so the choice of parameterisation is not a detail.
We could also not improve performance by modifying the \ac{GP} hyperpriors (see Appendix \ref{appendix:hypprior}).

\subsubsection{Marginal likelihood versus leave-one-out pseudo-likelihood}

It has been shown that the Bayesian marginal likelihood is not always the best likelihood objective for predictive uncertainty tasks \cite{skorna_2026_freeenergy}. 
A common variant is the \ac{LOO-CV} pseudo-likelihood, which can be substituted for the \ac{GP} likelihood and used for hyperparameter \ac{UQ} in the same manner.
A simple interpretation of the improved performance of the \ac{LOO-CV} pseudo-likelihood for some regression tasks is that the marginal likelihood evaluates the probability of the observed data under the assumed probabilistic model and can therefore be sensitive to model misspecification. 
In contrast, \ac{LOO-CV} directly evaluates out-of-sample predictive performance and does not require the probabilistic model to provide an accurate generative description of the data.

\begin{figure}
    \centering
    \includegraphics[width=\columnwidth]{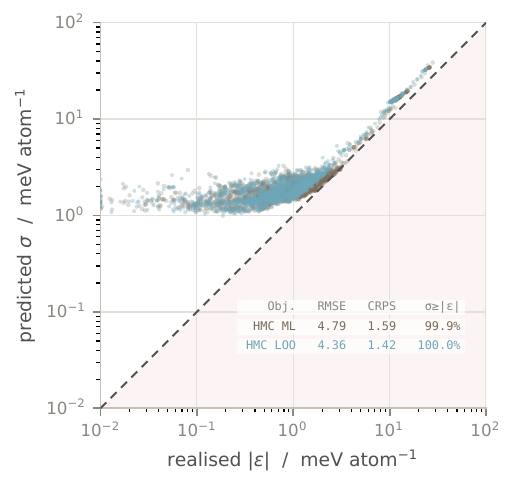}
    \caption{Comparison between \ac{ML} and \ac{LOO-CV} likelihood objective functions.}
    \label{fig:loo}
\end{figure}

Here we tested our physics-informed \ac{GP} equipped with the nonstationary Mat\'ern kernel and marginal likelihood/\ac{LOO-CV} pseudo-likelihood objectives and compared our evaluation metrics in Figure \ref{fig:loo}.
In this argon benchmark, the choice of likelihood objective made little difference to the \ac{RMSE}, \ac{CRPS}, and empirical coverage, with the \ac{LOO-CV} performing marginally better.
Here we used an energies only training set because the \ac{LOO-CV} pseudo-likelihood for \acp{GP} requires removing a single training example, which is complicated by the fact that a single configuration has one energy but multiple force calculations. 

\subsubsection{Homoskedastic versus heteroskedastic noise models}

We next examined whether allowing the observation noise to vary as a function of pair distance improves the predictive uncertainty of the hierarchical \ac{GP}.
The modelling choice of a fixed noise scale for all observations, known as a \textit{homoskedastic} noise model, was compared to a pair-distance-dependent noise term, a \textit{heteroskedastic} noise model, intended to allow larger uncertainty for configurations containing short-range or otherwise difficult pair environments. 
It has been shown that noise inference can significantly alter the \ac{UQ} capabilities of \ac{GP} models, with free energy \ac{UQ} specifically showing drastically improved calibration with a simple, fixed noise model inferred from data.
Both models used the same Mie prior mean, nonstationary Mat\'ern two-body kernel, energy and force training data, and HMC marginalisation over \ac{GP} hyperparameters.

In this benchmark, the heteroskedastic noise model slightly degraded predictive performance. 
For the 20-configuration energy+force-trained model, the homoskedastic \ac{GP} achieved an \ac{RMSE} of 1.02 meV atom$^{-1}$ and a \ac{CRPS} of 0.40 meV atom$^{-1}$, with empirical coverage of 99.9\%. 
The heteroskedastic model increased the \ac{RMSE} to 1.16 meV atom$^{-1}$ and the \ac{CRPS} to 0.42 meV atom$^{-1}$, while reducing coverage to 88.9\%.
Thus, in this low-data regime, the additional flexibility of the heteroskedastic noise model reduced the overdispersion of the predictive intervals, but at the cost of point accuracy and \ac{CRPS}.

This result suggests that the dominant uncertainty in the present \ac{GP} is not well described as input-dependent observational noise. 
The reference energies and forces are effectively deterministic electronic-structure calculations, so the residual uncertainty is probably epistemic model uncertainty arising from limited training data, imperfect kernel structure, model-form error, and hyperparameter uncertainty. 
Introducing a distance-dependent noise model may therefore absorb signal that would otherwise be explained by the \ac{GP} covariance, narrowing the predictive intervals at the expense of the point predictions.
For this reason, the homoskedastic noise model was retained as the default for \ac{GP} comparisons.

\begin{figure}
    \centering
    \includegraphics[width=\columnwidth]{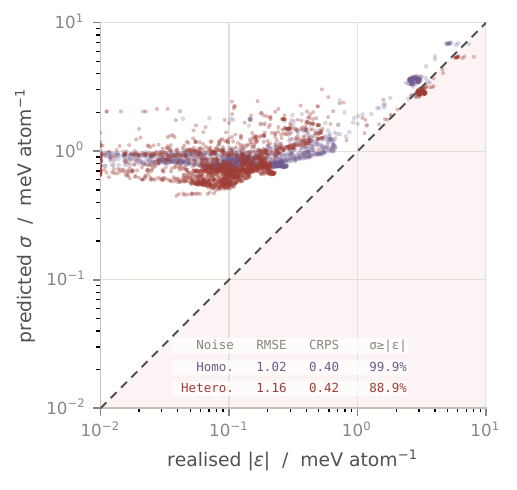}
    \caption{Comparison between homo- and heteroskedastic noise models.}
    \label{fig:noise}
\end{figure}

\subsection{Current landscape of \ac{GP}-based \acp{MLIP}}

Our controlled tests suggest that two features are particularly important for useful \ac{UQ} in GP-based \ac{MLIP}s: (1) an appropriate physics-informed prior mean, which produced the largest improvement in the low-data regime, and (2) propagation of hyperparameter uncertainty, which was particularly important for retaining uncertainty-error discrimination.
In contrast, the detailed covariance structure, likelihood objective, noise model, and informativeness of the hyperpriors produced comparatively small or mixed effects in this benchmark.
Reliable \ac{GP} uncertainty therefore does not follow automatically from the regression method itself, but depends on careful structural prior specification and treatment of uncertainties in the statistical model. 
This comes at a significant computational cost, particularly when the prior introduces additional hyperparameters that must themselves be inferred and propagated; however, lower-cost approximations such as the Laplace approximation may recover some of the benefits of hyperparameter marginalisation without requiring full \ac{HMC-NUTS} sampling.

Our results may also provide some insight into the behavior of ensemble-based \ac{MPNN} potentials. 
Independently initialized neural networks sample the outcomes of repeated stochastic optimisation rather than providing a well-defined posterior over potential energy surfaces. 
Consequently, there is no requirement that ensemble disagreement capture the full uncertainty of the model, and independently trained models can converge toward similar predictions even where systematic errors remain.
This provides one possible explanation for the near-zero ensemble variances observed for some MACE predictions in the present benchmark. 
Committees constructed from independently trained heads from different foundation models backbones probe a broader range of model variation and generally performed better here, although they similarly should not be interpreted as direct samples from a Bayesian posterior. 
Approximate Bayesian \ac{MPNN} methods may provide a more direct analogue to the hierarchical \ac{GP} approach and would be valuable for future comparison.

The movement from \acp{GP} toward \ac{MPNN} potentials has been driven by substantial advantages in scalability, computational efficiency, and representation of complex chemical environments. 
Nevertheless, recent developments in sparse and computationally efficient \ac{GP} methods \cite{noack_gp2scale_2025}, as well as multitask \cite{fisher_multitask_2024} and multi-observation \cite{yoowillow_active_2024} models, are addressing these limitations. 
We therefore expect \ac{GP}-based potentials to retain a complementary role where uncertainty itself is a primary quantity of interest, or necessary for uncertainty-critical applications where getting the answer wrong is risky. 
Active learning is one natural application, while multiscale modelling presents a more demanding case in which predictive uncertainty must ultimately be propagated into downstream quantities of interest. 
In both settings, the ability to explicitly represent and interrogate different sources of uncertainty remains a distinctive advantage of the \ac{GP} framework.

\section{\label{sec:discussion}Discussion}

The results presented here highlight a distinction that is often obscured in discussions of \ac{UQ} for \acp{MLIP}: predictive accuracy and useful uncertainty estimation are separate properties of a model. 
Across the argon benchmark, \ac{MACE} models equipped with standard \ac{UQ} protocols produced highly accurate energy predictions while their ensemble uncertainties substantially underestimated the observed errors and, in some cases, contained little information about which configurations were most difficult to predict. 
Conversely, the hierarchical \ac{GP} developed here generally avoided overconfidence and retained a positive relationship between predicted uncertainty and realised error, but produced predictive distributions that were substantially overdispersed relative to their nominal coverage in the low-realised error regime.
Neither behavior should therefore be interpreted simply as ``good'' or ``bad'' \ac{UQ}.
Rather, these results expose several distinct properties that must be considered when assessing the usefulness of an uncertainty model.

At least three aspects of predictive uncertainty are relevant in this context. 
First, a probabilistic model should ideally be ``calibrated'', such that intervals assigned a given probability contain the corresponding fraction of observations. 
Second, the predictive distributions should be sufficiently ``sharp'' to provide useful information; trivially broad intervals can achieve high empirical coverage without providing useful constraints on the prediction.
Third, the magnitude of the predicted uncertainty may be expected to ``discriminate between relatively reliable and unreliable predictions'', particularly when uncertainty is used for active learning or extrapolation detection. 
Proper scoring rules such as \ac{CRPS} combine aspects of the first two criteria, whereas correlations between predicted uncertainty and realised error primarily probe the third. 
The present benchmark demonstrates that these properties need not improve together.

This distinction is particularly apparent for the hierarchical \ac{GP}. 
Its nominal one-standard-deviation intervals frequently contain substantially more than the expected 68\% of test observations and therefore should not be described as calibrated in an absolute probabilistic sense. 
Instead, the model is systematically conservative or overdispersed.
The opposite failure mode is observed for several ensemble models, where accurate mean predictions are accompanied by narrow uncertainty estimates that can substantially undercover the observed errors and, in some cases, fail to rank difficult configurations reliably.

Which of these properties matters most depends on the intended use of the potential. 
In active learning, the absolute probabilistic interpretation of an uncertainty estimate may be less important than its ability to rank configurations according to where additional electronic-structure calculations would be most informative. 
An uncertainty estimator that is systematically too large but preserves this ordering could remain useful for selecting new configurations. 
By contrast, an uncertainty estimate that is sharp but poorly correlated with model failure could cause an active-learning procedure to overlook precisely those configurations for which additional training data are required. 
For uncertainty propagation through multiscale simulations, however, ranking alone is insufficient. 
In that setting, uncertainty is assigned a quantitative probabilistic meaning and propagated into downstream quantities of interest, making both calibration and sharpness important. 
Severe overdispersion would then propagate unnecessarily large uncertainties, while underdispersion could produce unjustifiably precise downstream predictions.

These considerations also raise the more fundamental question of what an \ac{MLIP} uncertainty estimate is intended to represent. 
The electronic-structure reference calculations considered here are effectively deterministic, such that the dominant uncertainty is not conventional measurement noise. 
Instead, uncertainty arises from incomplete knowledge of the potential energy surface due to finite training data, limitations of the chosen representation and covariance structure, and uncertainty in the parameters defining the regression model. 
These contributions are often collectively described as epistemic or model uncertainty, but they are not necessarily equivalent. 
For example, the conditional \ac{GP} variance describes uncertainty within a specified \ac{GP} model at fixed hyperparameters, whereas marginalisation over the hyperposterior additionally represents uncertainty concerning the appropriate mean and covariance functions. 
Neither contribution necessarily captures errors arising because the assumed model class itself is inadequate. 
Consequently, interpreting a single predicted standard deviation as a complete measure of ``the error'' of an \ac{MLIP} risks conflating several conceptually different sources of uncertainty.

The comparison with \ac{MACE} should be interpreted within the deliberately restricted scope of this benchmark. 
The \ac{GP} model has been constructed specifically around the simple physics of argon, for which an informative analytical two-body prior can be formulated and the relevant configurational space is low dimensional.
Modern \ac{MPNN} potentials are designed for considerably more heterogeneous chemical environments, where constructing equally informative \ac{GP} priors may be substantially more difficult. 
The present results therefore do not establish that \ac{GP}-based potentials generally provide superior \ac{UQ} to neural-network potentials.
Instead, they demonstrate that the particular ensemble strategies examined here can provide poor proxies for predictive uncertainty even when their underlying models are highly accurate. 

More generally, these results suggest that there may be no universally optimal definition of \ac{MLIP} uncertainty independent of its intended application. 
A model intended for active learning may prioritise discrimination and reliable identification of extrapolative configurations; whereas a model used for uncertainty propagation may require quantitatively calibrated predictive distributions; and a model intended primarily for high-throughput prediction may place substantially greater weight on sharpness and computational efficiency. 
Evaluating \ac{UQ} methods using a single metric can therefore conceal important failure modes. 
We suggest that \ac{MLIP} uncertainty models should instead be assessed jointly in terms of predictive accuracy, probabilistic calibration, sharpness or proper scoring rules, and their ability to discriminate between predictions of differing reliability.

Within this framework, the principal advantage of hierarchical \ac{GP} models is markedly not that they provide perfectly calibrated uncertainty. 
Indeed, the present results demonstrate explicitly that they do not. 
Rather, their value lies in providing a probabilistically structured model in which different assumptions and sources of uncertainty can be exposed, modified, and propagated systematically. 
The remaining challenge then is to retain this structure while obtaining posterior predictive distributions that are both sufficiently conservative to avoid confidently incorrect predictions and sufficiently sharp to be useful in downstream simulations. 
Resolving that tradeoff, and determining which aspects of uncertainty must be represented for different \ac{MLIP} applications, remains an important open problem for uncertainty-aware interatomic potential development.

\section{\label{sec:conclusions}Conclusions}

Accurate interatomic potentials do not necessarily provide useful uncertainty estimates. 
In this controlled argon benchmark, \ac{MACE} models achieved excellent predictive accuracy, while uncertainties obtained from several ensemble strategies substantially underestimated realised errors and, in some cases, lost the ability to identify configurations associated with larger prediction errors. 
In contrast, hierarchical \ac{GP}-based models generally retained a positive relationship between predicted uncertainty and realised error and avoided severe overconfidence, although were often too conservative for lower error test configurations.

Our analysis of the \ac{GP} model identifies two features that are particularly important in the present benchmark: an appropriate physics-informed prior mean and propagation of uncertainty in the \ac{GP} hyperparameters. 
The prior mean produced the largest improvement in the low-data regime, while hyperparameter marginalisation was particularly important for retaining uncertainty-error discrimination.
Notably, a Laplace approximation retained much of the error-discrimination behavior obtained by full hyperposterior sampling, suggesting a potentially lower-cost route to hyperparameter \ac{UQ}.
By comparison, changing the likelihood objective, noise model, or imposing increasingly restrictive hyperpriors provided little additional benefit in our benchmark. 
Importantly, however, full hyperparameter marginalization did not automatically produce calibrated uncertainty, demonstrating that a formally Bayesian treatment alone is insufficient to guarantee useful predictive distributions.

We emphasize that ``good'' \ac{UQ} cannot be characterized by a single metric. 
Calibration, sharpness, error discrimination, and predictive accuracy represent distinct properties whose relative importance depends on the intended application. 
Establishing how these properties should be balanced, and determining which sources of uncertainty must be represented for active learning, uncertainty propagation, and other uncertainty-critical applications, remains a central challenge for the development of trustworthy \acp{MLIP}.

\begin{acknowledgments}

This work was financially supported by a Leverhulme Trust Research Project Grant (RPG-2017-191), the Engineering and Physical Science Research Council (EPSRC) under grant EP/R043612/1, the CASTEP-USER project funded by UK Research and Innovation under the grant agreement EP/W030438/1, and the NOMAD Centre of Excellence (European Commission grant agreement ID 951786). GS is supported by a studentship from the EPSRC Centre for Doctoral Training in Modelling of Heterogeneous Systems (HetSys II) funded by the EPSRC under grant EP/Y035429/1. We acknowledge computational resources provided by the Scientific Computing Research Technology Platform of the University of Warwick, STFC Scientific Computing Department’s SCARF cluster, the EPSRC-funded HPC Midlands+ consortium (EP/P020232/1, EP/T022108/1) and on ARCHER2 (https://www.archer2.ac.uk/) via the UK Car-Parinello consortium (EP/P022065/1). BLS acknowledges funding from the Institute of Organic Chemistry and Biochemistry postdoctoral fellowship program. For the purpose of Open Access, the author has applied a CC-BY public copyright licence to any Author Accepted Manuscript (AAM) version arising from this submission.

\end{acknowledgments}

\section{Appendix}

\subsection{Evaluation metrics}\label{appendix:eval}

\paragraph{Root mean square error (RMSE)}: \ac{RMSE} is a measure of distance between a predicted $\hat{\mathbf{y}}$ and observed $\mathbf{y}$ given by $\text{RMSE} := \sqrt{\frac{1}{n} \sum_{i=1}^n ||\hat{\mathbf{y}}_i - \mathbf{y}_i||^2}$, where $n$ is the total number of observation vectors.
\ac{RMSE} is sensitive to outliers due to the square penalty term before averaging.


\textit{Parity plots:} A parity plot compares the realised absolute prediction error, $|\epsilon|$, with the predicted standard deviation, $\sigma$. 
The line $\sigma = |\epsilon|$ provides a reference for whether the predicted uncertainty is larger or smaller than the realised error for an individual prediction, but should not be interpreted as the expected pointwise relationship for a calibrated probabilistic model. 
Correlation between $\sigma$ and $|\epsilon|$ instead provides a measure of the ability of the uncertainty estimate to discriminate between predictions with relatively small and large realised errors.

\subsection{Hyperpriors}\label{appendix:hypprior}

\paragraph{Default hyperpriors} For the hierarchical \ac{GP} models, weakly informative hyperpriors were assigned to the Mie prior-mean parameters, residual kernel parameters, and observation-noise scales.
The Mie parameters were centered on physically reasonable argon values, while retaining broad support so that the data could adjust the well depth, size parameter, and repulsive exponent.
The residual \ac{GP} kernel hyperpriors were chosen to allow order-meV energy corrections over Angstrom-scale variations in pair distance, with additional floor and sparse-distance terms controlling the nonstationary Gibbs-Mat\'ern covariance structure.
Separate energy and force noise scales were included for joint energy-force training.
The default hyperpriors used in this work are:
\[
\begin{aligned}
\log \epsilon &\sim \mathcal{N}(\log 0.01,\,1.0^2),\\
\log \sigma_\mathrm{Mie} &\sim \mathcal{N}(\log 3.4,\,0.2^2),\\
\lambda &\sim \mathcal{N}(12,\,3.0^2)\ \mathrm{truncated\ to}\ [6.01,30],\\
\log \ell_2 &\sim \mathcal{N}(\log 1.0,\,1.0^2),\\
\log A_2 &\sim \mathcal{N}(\log 0.01,\,2.0^2),\\
\log \gamma_2 &\sim \mathcal{N}(\log 0.37,\,1.0^2),\\
r_0 &\sim \mathcal{N}(3.8,\,1.5^2)\ \mathrm{truncated\ to}\ [1,10],\\
\log \kappa_2 &\sim \mathcal{N}(\log 4.5,\,1.0^2),\\
\log \delta_{\mathrm{floor}} &\sim \mathcal{N}(\log 5\times10^{-4},\,1.0^2),\\
\log \delta_{\mathrm{sparse}} &\sim \mathcal{N}(\log 3.7\times10^{-3},\,1.5^2),\\
\log \ell_{\mathrm{sparse}} &\sim \mathcal{N}(\log 0.6,\,1.0^2),\\
\log \sigma_E &\sim \mathcal{N}(\log 10^{-3},\,1.0^2),\\
\log \sigma_F &\sim \mathcal{N}(\log 10^{-3},\,1.5^2).
\end{aligned}
\]

\paragraph{Hyperprior sensitivity analysis} Changing the hyperpriors of the physics-informed prior ($\lambda$-6) Mie potential mean is a possible route to affect the conservative predictions of the \ac{GP}-based \ac{MLIP}.
Here we tested our default hyperprior against a series of increasingly informative hyperpriors summarized in Table \ref{tab:mie_hyperpriors}.
The default model used broad physics-informed priors centered near typical Ar parameters, while three additional models imposed increasingly restrictive priors around $\epsilon = 0.24$ kcal mol$^{-1}$, $\sigma = 3.4$ \AA, and $\lambda = 12$. 
All models used the same Mie prior mean, nonstationary Mat\'ern kernel, homoskedastic noise model, energy+force training data, and hyperparameter marginalisation.

Tightening the Mie hyperpriors only marginally improved performance. 
The default hyperpriors gave an \ac{RMSE} = 1.10 meV atom$^{-1}$, \ac{CRPS} = 0.44 meV atom$^{-1}$, and empirical one-sigma coverage of 99.2\%. 
The increasingly restrictive Ar-centered priors produced marginally smaller \ac{RMSE} values of 0.91, 0.94, and 1.02 meV atom$^{-1}$ for the loose, medium, and strict settings, respectively. 
\ac{CRPS} exhibited the same trend with values of 0.35, 0.37, and 0.40 meV atom$^{-1}$, while coverage decreased to approximately 88-89\% in the medium and strict hyperprior cases.
Overall, tightening the Mie hyperpriors produced modest improvements in \ac{RMSE} and \ac{CRPS}, while the medium and strict hyperpriors substantially reduced overcoverage.

These results show that stronger physical prior information is not automatically beneficial, and only marginally affects \ac{UQ} metrics in this training regime. 
Although the tighter priors encode reasonable isolated Ar$_2$ parameters, the \ac{GP} is trained on a mixed set of dimer and trimer configurations with finite-data uncertainty and residual many-body effects. 
Overly restrictive Mie priors can therefore limit the ability of the hierarchical model to adapt the prior mean and covariance jointly to the training distribution. 
The broader default hyperpriors provide enough physical structure to regularize the low-data problem while still allowing the posterior to adjust to the observed configurations.

\begin{table*}
\centering
\caption{Mie prior-mean hyperpriors tested in the hyperprior sensitivity study. 
All cases use $\sigma_0 = 3.4$ \AA, $\lambda_0 = 12$, and, for the Ar-centered cases, $\epsilon_0 = 0.24$ kcal mol$^{-1}$ = 0.0104 eV.}
\label{tab:mie_hyperpriors}
\begin{tabular}{llll}
\toprule
Case & $\sigma$ prior & $\epsilon$ prior & $\lambda$ prior \\
\midrule
Default 
& $\log\sigma \sim \mathcal{N}(\log 3.4, 0.20^2)$ 
& $\log\epsilon \sim \mathcal{N}(\log 0.01, 1.00^2)$ 
& $\lambda \sim \mathcal{N}(12, 3.00^2)$ \\

Loose 
& $\log\sigma \sim \mathcal{N}(\log 3.4, 0.10^2)$ 
& $\log\epsilon \sim \mathcal{N}(\log 0.0104, 0.50^2)$ 
& $\lambda \sim \mathcal{N}(12, 1.50^2)$ \\

Medium 
& $\log\sigma \sim \mathcal{N}(\log 3.4, 0.05^2)$ 
& $\log\epsilon \sim \mathcal{N}(\log 0.0104, 0.25^2)$ 
& $\lambda \sim \mathcal{N}(12, 0.75^2)$ \\

Strict 
& $\log\sigma \sim \mathcal{N}(\log 3.4, 0.02^2)$ 
& $\log\epsilon \sim \mathcal{N}(\log 0.0104, 0.10^2)$ 
& $\lambda \sim \mathcal{N}(12, 0.30^2)$ \\
\bottomrule
\end{tabular}
\end{table*}

\begin{figure}
    \centering
    \includegraphics[width=\columnwidth]{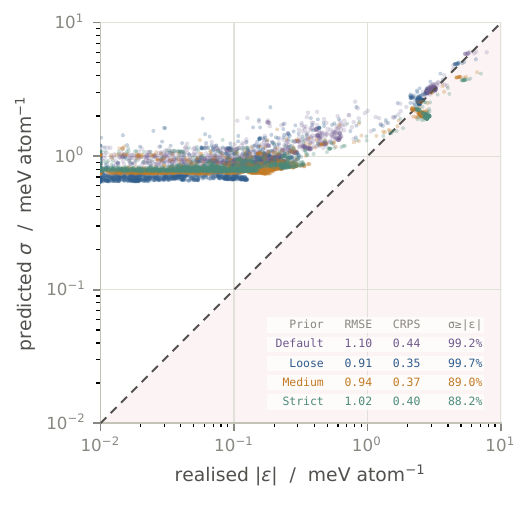}
    \caption{Analysis of increasingly informative hyperpriors.}
    \label{fig:hyperpriors}
\end{figure}

\section*{References}

\bibliography{refs}

\end{document}